\documentclass[twocolumn,aps,pre,floatfix,superscriptaddress]{revtex4-1}
\usepackage{graphics}
\usepackage{graphicx}
\usepackage{amsmath}
\usepackage{amssymb}
\usepackage{ulem}
\usepackage{color}

\begin{document}
%\bibliographystyle{plain}
%\pagestyle{plain}
%\pagenumbering{arabic}%\rmfamily
\title{Dynamical Scaling Implications of Ferrari, Pr\"{a}hofer, and Spohn's Remarkable Spatial Scaling Results for Facet-Edge Fluctuations}
\author{T.L. Einstein}
\email[]{einstein@umd.edu}
%\ead[url]{http://www2.physics.umd.edu/~einstein/}
\affiliation{Department of Physics, University of Maryland, College Park, Maryland 20742-4111, USA}
\affiliation{Condensed Matter Theory Center, University of Maryland, College Park, Maryland 20742-4111, USA}
\author{Alberto Pimpinelli}
\email[]{ap19@rice.edu}
\affiliation{Department of Physics, University of Maryland, College Park, Maryland 20742-4111, USA}
\affiliation{Rice Quantum Institute, Rice University, Houston, Texas 77005, USA}
\date{\today}
% Article starts here

\begin{abstract}
Spurred by theoretical predictions from Spohn and coworkers [Phys. Rev. E {\bf 69}, 035102(R) (2004)], we rederived and extended their result heuristically as well as investigated the scaling properties of the associated Langevin equation in curved geometry with an asymmetric potential.  With experimental colleagues we used STM line scans to corroborate their prediction that the fluctuations of the step bounding a facet exhibit scaling properties distinct from those of isolated steps or steps on vicinal surfaces.  The correlation functions was shown to go as $t^{0.15(3)}$ decidedly different from the $t^{0.26(2)}$ behavior for fluctuations of isolated steps.  From the exponents, we were able to categorize the universality, confirming the prediction that the non-linear term of the KPZ equation, long known to play a central role in non-equilibrium phenomena, can also arise from the curvature
or potential-asymmetry contribution to the step free energy. We also considered, with modest Monte Carlo simulations, a toy model to show that confinement of a step by another nearby step can modify as predicted the scaling exponents of the step's fluctuations.
This paper is an expansion of a celebratory talk at the 95$^{\rm th}$ Rutgers Statistical Mechanics Conference, May 2006.
\end{abstract}

\maketitle

\section{Introduction}
    %Technological demands on the fabrication and properties of nano-structures \cite{3} provide renewed motivation %for understanding the properties that control morphology changes on the nanoscale.

Fluctuations of steps on surfaces play a central role in determining their impact on surface processes and the evolution of surface morphology.  In the past nearly-two decades, the step continuum model has allowed several successful quantitative correlations of direct observations of step fluctuations with kinetic and thermodynamic descriptions of nanoscale structural evolution \cite{JW,giesen01,LDEW,6,6+}, bridging from the atomistic and nanoscale to the mesoscale.   For steps on flat or vicinal (misoriented modestly from a facet orientation) surfaces, there are two well-defined scaling behaviors for temporal correlations, corresponding to cases B and A, conserved and non-conserved dynamics, respectively, in the framework of dynamic critical phenomena \cite{HH}. Several examples of both behaviors have been observed experimentally in physical systems \cite{JW,giesen01,tle-Dux} and numerically in Monte Carlo simulations \cite{measuring,NCBBrownian,SelkeB}.

For complex structures where mass transport is limited by geometry, the fundamental question of how fluctuations behave in a constrained environment becomes experimentally accessible.  These issues become particularly important for smaller structures, especially nanostructures, where issues of finite volume (shape effects and volume conservation) become non-negligible \cite{KCST,8}.  Although the step can still be viewed as a 1D interface obeying a Langevin-type equation of motion, not only local deformation but global effects must be considered when calculating the step chemical potential.  These considerations alter the equation of motion, including the noise term, resulting in different \textcolor[rgb]{0.00,0.00,0.00}{universality} classes of dynamic scaling \cite{9} \textcolor[rgb]{0.00,0.00,0.00}{(see Table II below)}.

\begin{figure}[b]
\includegraphics[width=8 cm]{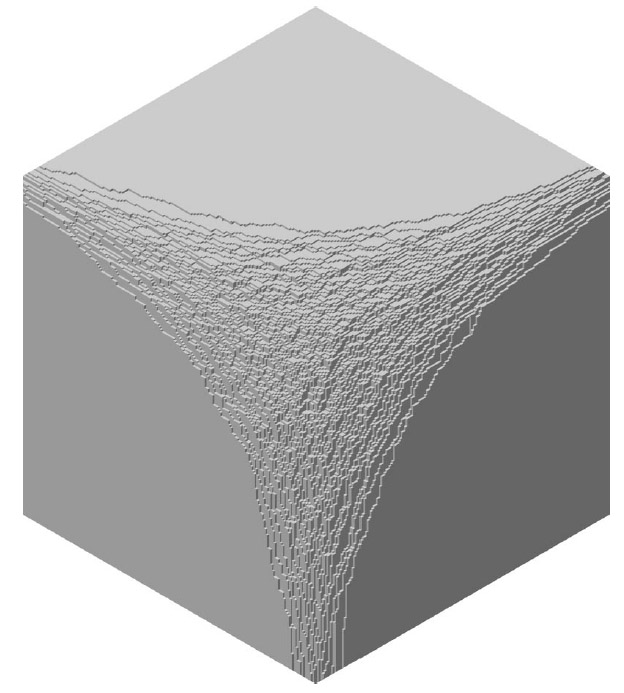}
\caption {Simple-cubic crystal corner viewed from the \{111\} direction, from Refs.~\cite{ferrari04,FS,Fphd,Spohn06}.}
{\label{f:FPSCorner}}
\end{figure}

Thus, it was especially enlightening and inspiring to read of a well-defined intermediate scaling regime in Ferrari, Pr\"{a}hofer, and Spohn's stimulating paper \cite{ferrari04} (hereafter FPS) (as well as related works \cite{FS,Fphd,Spohn06}), in which they computed the scaling of equilibrium fluctuations of an atomic ledge bordering a
crystalline facet surrounded by rough regions of the equilibrium crystal shape in their examination of a 3D Ising corner (Fig.~\ref{f:FPSCorner}).
 We refer to this boundary edge as the ``shoreline" since it is the edge of an island-like region--the crystal facet--surrounded by a ``sea" of steps.

  FPS derive an intriguing exact result, concerning how
the width $w$ of shoreline fluctuations scales as
a function of the linear size of the facet. This length corresponds to the length of a step or the linear dimension of an island (or its circumference).  This length is often called $L$ \cite{pimpi,D-PRL} and other times $\ell$ \cite{P-SSL} (while $\ell$ has a closely related but slightly different meaning in Ref.~\cite{ferrari04}).  To prevent any possible confusion, we denote this length by the Polish crossed $L$, \textit{\L}, in this paper, following the notation used in a ceremonial presentation on this subject \cite{rutgers}.  We anticipate that
\begin{equation}
w\sim \mbox{\textit{\L}}^\alpha,
\label{e:wL}
\end{equation}
\noindent where the value of roughness exponent $\alpha$ depends on the mode of mass transport and the geometry of the step.  For the step that serves as the border a two-dimensional (2D) island on a high-symmetry crystal plane, one expects (and finds in physical and numerical experiments) that $w\sim \mbox{\textit{\L}}^{1/2}$, i.e.\ $\alpha = 1/2$, since this step performs a random walk \cite{BisaniSelke}.

FPS show that, as we quipped in the title of our paper \cite{P-SSL}, ``a crystal facet is not an island". Indeed, they find that instead of the expected random-walk behavior,
\begin{equation}
w\sim \mbox{\textit{\L}}^{1/3},
\label{e:wL3rd}
\end{equation}
\noindent i.e.\ $\alpha = 1/3$, for a crystal facet. They prove that the origin of the unusual $\mbox{\textit{\L}}^{1/3}$ scaling lies in the step-step interactions between the facet
ledge and the neighboring steps under conditions of conserved volume.  Note that this value of $\alpha$ is intermediate between $\alpha = 1/2$ for isolated steps and $\alpha = 0$ ($w\sim \mbox{ln(\textit{\L})})$) \cite{role} for a step on a vicinal surface, i.e.\ in a step train.

FPS's formidable calculation is based on the use of
free spinless fermions, transfer
matrices, random-matrix properties, Airy functions, and specific
models;
as a purely static result, it does not
address the question of the time behavior of step fluctuations, which are easier to measure experimentally.

This article is an expansion of a celebratory talk \cite{rutgers} which described the impact of FPS on our research, in particular the results found in three publications \cite{P-SSL,D-PRL,DSPEW}.  In Section II we summarize highlights of FPS that motivated and underpinned our subsequent work.  In Section III we describe the relevant correlation functions.  Next we present a heuristic derivation extending the reasoning of Pimpinelli et al. \cite{pimpi} that leads to the dynamic scaling of shoreline fluctuations, as well as the static result of FPS.  Then we present a more formal analysis of scaling for curved steps in an asymmetric potential.  In Section IV we describe experiments using scanning tunneling microscopy (STM) that demonstrate the novel scaling behavior in a physical system.  In Section V we present Monte Carlo results for a toy model that shows in a simple system the effect of a neighboring step on the fluctuations of a step.  The Conclusion section offers some final remarks.

\begin{figure}[t]
\includegraphics[width=8 cm]{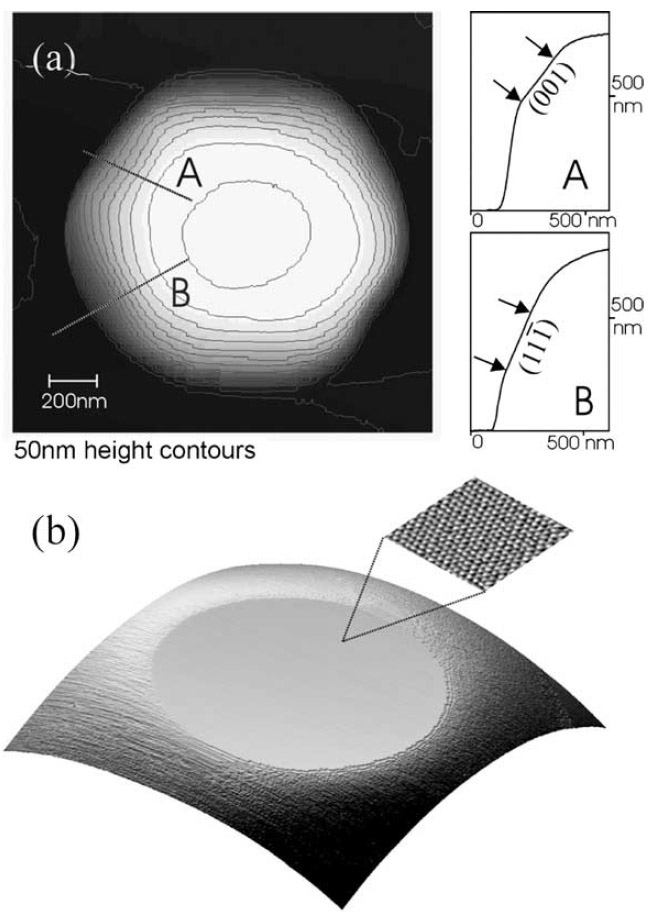}
\caption {(a) Micron-size lead crystal (supported on Ru) imaged with a variable-temperature
STM at $T = 95^{\circ}$C. Annealing at $T = 95^{\circ}$C for 20 hours allowed it to obtain its stable, regular shape. Lines marked A and
B indicate location of profiles. Profile A crosses a (0 0 1)-side
facet, while profile B a (1 1 1)-side facet. (b) 770 nm$\times$770 nm
section of the top part of a Pb-crystal. The insert shows a 5.3
nm$\times$5.3 nm area of the top facet, confirming its (1 1 1)-orientation.
Both the main image and the insert were obtained at $T = 110^{\circ}$C; from Ref.~\cite{Thurmer03}.
}
{\label{f:ThurmerPb}}
\end{figure}
\begin{figure}[b]
\includegraphics[width=8 cm]{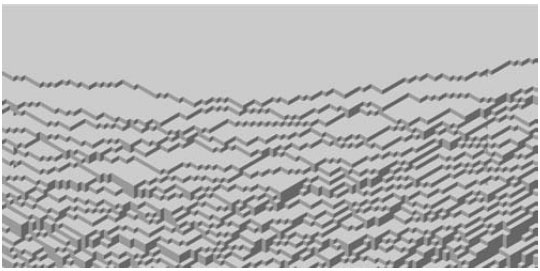}
\includegraphics[width=8 cm]{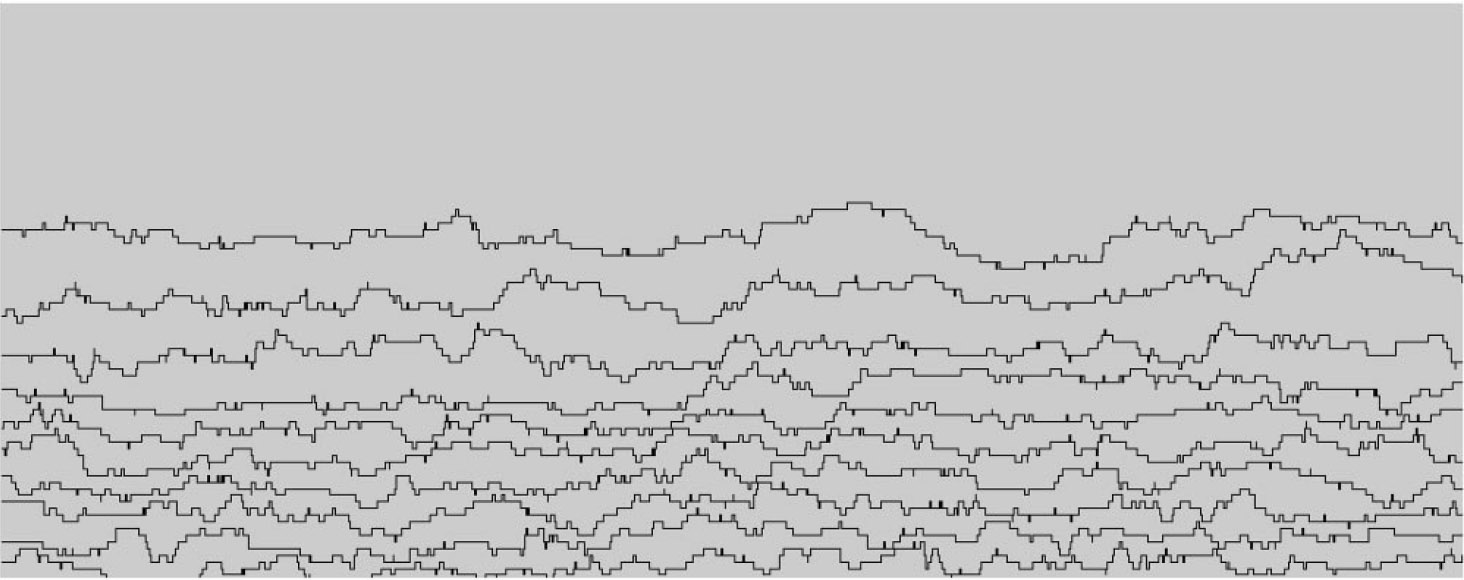}
\caption {a)  Magnified detail of the steps near the facet edge in Fig.~\ref{f:FPSCorner}, from Refs.~\cite{FS,Fphd};  b) Snapshot of computed configurations of the top steps (those near a facet at the flattened side portion of a cylinder) for a terrace-step-kink (TSK) model with volume constraint, from Refs.~\cite{ferrari04,Fphd}.
}
{\label{f:FPSSteps}}
\end{figure}

\section{Summary of Highlights of FPS}

FPS assume that there are no interactions between steps beyond entropic (i.e., the steric repulsions arising from the fact that steps cannot cross), so that the step configurations can be mapped to the world lines of free spinless fermions; the entropic repulsion is captured by the fermionic Pauli condition \cite{touch}.  A key new feature compared to treatments of vicinal surfaces is that the volume of the crystallite is conserved. Their fermionic Hamiltonian $\mathcal{H}_F$ is

\begin{equation}
\mathcal{H}_F = \sum_{j} \left( -a_j^\dagger a_{j\! +\! 1} -a_{j\! +\! 1}^\dagger a_j +2a_j^\dagger a_j + \frac{j}{\lambda}a_j^\dagger a_j   \right),
\label{e:HF}
\end{equation}
where $\lambda^{-1}$ is a Lagrange multiplier associated with conserved volume. See Fig.~\ref{f:FPSSteps}. It is this final term that is new in their treatment.  Its asymmetry is key to the novel behavior they find.  They then derive an exact result for the step density  in terms of $J_j$, the Bessel function of integer order $j$, and its derivative.  Near the shoreline they find
\begin{equation}
\lim_{\lambda \rightarrow \infty} \lambda^{1/3} \rho_\lambda (\lambda^{1/3}x) = -x (\mathrm{Ai}(x))^2 +(\mathrm{Ai}^\prime(x))^2,
\end{equation}
\noindent \textcolor[rgb]{0.00,0.00,0.00}{where $\rho_\lambda$ is the step density (for the particular value of $\lambda$).}

The presence of the Airy function Ai results from the asymmetric potential implicit in $\mathcal{H}_F$ and preordains exponents involving 1/3.  The variance of the wandering of the shoreline, the top fermionic world line in Fig.~\ref{f:FPSSteps} and denoted by $b$, is given by
\begin{equation}
\mathrm{Var}[b_\lambda(t) - b_\lambda(0)] \cong \lambda^{2/3} g(\lambda^{-2/3}t)
\label{e:var1}
\end{equation}
\noindent where \textcolor[rgb]{0.00,0.00,0.00}{$t$ is the fermionic ``time" along the step;} $g(s) \sim 2|s|$ for small $s$ (diffusive meandering) and $\sim 1.6264 - 2/s^2$ for large $s$.  They then \textcolor[rgb]{0.00,0.00,0.00}{set $\lambda$ to} a scaling parameter $\ell = (4N/1.202 \ldots)^{1/3}$, where $1.202 \ldots$ is Apery's constant and $N$ is the number  of atoms in the crystal, as in Fig.~\ref{f:FPSCorner}.  They find
\begin{equation}
\mathrm{Var}[b_\ell(\ell \tau + x) - b_\ell(\ell \tau)] \cong ({\cal A}\ell)^{2/3} g\left({\cal A}^{1/3}\ell^{-2/3}x\right),
\label{e:var2}
\end{equation}
where ${\cal A} = (1/2)b_\infty^{\prime\prime}$ \cite{calA}.  This leads to the central result that the width $w \sim \ell^{1/3}$.  Furthermore, the fluctuations are non-Gaussian.  They also show that near the shoreline, the deviation of the equilibrium crystal shape from the facet plane takes \textcolor[rgb]{0.00,0.00,0.00}{on the Gruber-Mullins-Pokrovsky-Talapov \cite{GMPT} form} \mbox{$-(r-r_0)^{\textcolor[rgb]{0.00,0.00,0.00}{3/2}}$}, where $r$ is the lateral distance from the facet center and $r_0$ is the radius of the facet.

\section{Analytical Results}

In this section, we discuss computation of the time scaling of step-edge fluctuations
using two non-rigorous approaches.
First, we adopt a simple scaling
argument, starting from FPS's exact result. Then we derive a
continuum-equation description of the step
bordering a crystal facet. Then, with simple power counting we
rederive FPS's result, as well as the temporal power-law scaling of edge fluctuations.

For straight steps, which underlie treatments of this problem, one adopts
cartesian coordinates $(x,y)$, $y$ being in the direction along
the step edge, and
$x(y)$ describing the step profile, in what has been called ``Maryland notation" \cite{MDnotn}. We focus attention on the step autocorrelation function

\begin{equation}
G(t) = \langle[x(y_0,t+t_0) -x(y_0,t_0)]^2\rangle_{y_0,t_0} \; \raisebox{-1.5ex}{$\stackrel{\textstyle \sim}{\scriptstyle t\rightarrow 0}$}\; t^{2\beta},
\label{e:Gt}
\end{equation}
\noindent which can readily be computed in a Monte Carlo simulation \cite{measuring} and measured experimentally with a scanning probe like STM.  It is less feasible to measure spatial correlation functions since such experiments do not take an instantaneous ``snapshot."  Like a television screen, different parts of the micrograph correspond to different times, and it is problematic to deal with what transpires between successive visits by the STM tip to nearby positions.  Furthermore, in such experiments one does not do a full average over $y_0$ but rather picks a single value; for that case we replace $G(t)$ by $G(y_0,t)$, for which there is no spatial average.  The resulting plot of displacement $x$ vs.\ time looks similar to scans of $x$ along a step, and so are called ``pseudo-images" \cite{giesen01}. (Cf.\ Fig.~\ref{f:linescan} below.)  At short times
$G(y_0,t)$ exhibits the same $t^{2\beta}$ behavior as $G(t)$.  The value of exponent $\beta$ depends on the
atomistic processes responsible for the fluctuations of the step, but
also on the position of the step
with respect to a crystal facet, as we proved in Ref.~\cite{P-SSL} and recount shortly below. In the other limit, $G(t\rightarrow \infty)$ saturates to $2w^2$, where $w$ is width of the fluctuations.

When doing simulations \cite{role} (or if one had a probe that could take instantaneous snapshots), one can probe
\begin{equation}
G(y,t_0) = \langle[x(y_0+y,t_0) -x(y_0,t_0)]^2\rangle_{y_0} \sim w^2,
\label{e:Gy}
\end{equation}
\noindent for large values of $y$.
Then the roughness exponent $\alpha$ can be extracted from the saturation value of the width $w$ of the fluctuating step by using Eq.~(\ref{e:wL}) and identifying \textit{\L} with the size of the system in the $y$-direction.

\subsection{Heuristic Derivation}
\label{s:heurist}

\begin{figure}[t]
\includegraphics[width=8 cm]{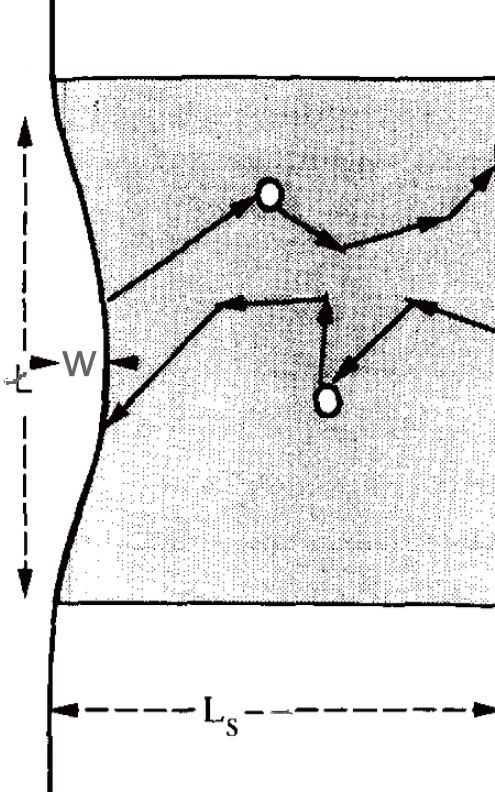}
\caption {Schematic of a stepped surface seen from above, to illustrate the ``pipe" concept of Ref.~\cite{pimpi}.  \textcolor[rgb]{0.00,0.00,0.00}{The fluctuating step, depicted as a thick wavy line, shows the
definitions of \textit{w} and \textit{\L}. For the two particular cases considered in the text, the thickness $L_s$ of the pipe-like reservoir to the right of the step, in which atom exchange occurs, is of order the lattice spacing $a$.}  Adapted from Ref.~\cite{pimpi}.
}
{\label{f:PVpipe}}
\end{figure}

Starting from FPS's  result that the roughness exponent $\alpha = 1/3$ in Eq.~(\ref{e:wL}),
we apply Pimpinelli {\it et al.}'s argument \cite{pimpi} for finding the temporal
scaling:  Referring to Fig.~\ref{f:PVpipe}, consider a \textcolor[rgb]{0.00,0.00,0.00}{protruding portion of step of length $\mbox{\textit{\L}}$ and width $w$. At equilibrium, the protrusion is due to fluctuations in the number of atoms to and from a``reservoirÓ. The reservoir is assumed to be, depending on the limiting atom transport process,
either another part of the same step, or a region on the crystal surface. Before considering a specific kinetic process, let us derive a few general relations.}  \textcolor[rgb]{0.00,0.00,0.00}{On average, a number $N(t)$ of atoms continually move between the step and the reservoir during a time interval
$t$. The reservoir is by definition situated one diffusion length $L_s$ away from the protrusion.}
While the net flux \textcolor[rgb]{0.00,0.00,0.00}{to and from the step vanishes, the
 number $N(t)$ }fluctuates
around its vanishing mean; \textcolor[rgb]{0.00,0.00,0.00}{assuming that atom fluxes in different time intervals are uncorrelated, we can compute}
the typical size of the fluctuation $\delta N$, which is  of order $\sqrt
{N(t)}$. Denoting by $\Omega$ the atomic
area, we
estimate the size $w$ of the protrusion along the step edge (defined, as said above, as the
amplitude or width of a typical step
fluctuation of length
$\mbox{\textit{\L}}$) from $w\times \mbox{\textit{\L}}\approx\Omega\sqrt {N(t)}$. To estimate $N(t)$, we note that the \textcolor[rgb]{0.00,0.00,0.00}{size (surface area) of the reservoir} feeding the
fluctuation  is $\mbox{\textit{\L}} L_s$. Then the number of atoms moving to and from the step edge during
time $t$ is proportional to the
number of diffusing atoms in the
region feeding the step, $c_{eq}\mbox{\textit{\L}} L_s$ (where $c_{eq}$ is the
equilibrium particle
density) and to the fraction of time
the atoms spend in this region, $t/
\tau^*$. The
characteristic time $\tau^*$ depends on the specific transport process
(see below). Then, as in Ref.~\cite{pimpi}:
\begin{equation}
\label{N}
N(t)\approx\frac {c_{eq}}{\tau^*}\mbox{\textit{\L}}L_s t.
\end{equation}
Furthermore, the squared area of the fluctuating bulge is
\begin{equation}
w^2\mbox{\textit{\L}}^2\approx (\delta N)^2\approx N,
\label{e:areafluc}
\end{equation}
where we assume $\delta N \approx \sqrt N$.

While Ref.~\cite{pimpi} considers several different cases, we focus here on the two primary scenarios: i) non-conservative mass
transport by attachment-detachment to/from the step edge, with fast
terrace diffusion (A/D) and ii) conservative mass transport by step-edge diffusion (SED).

i) In the A/D case,
${1}/{\tau^*}\approx k$, where $k$ is an appropriate kinetic coefficient.
For fast surface diffusion, the step
effectively
exchanges atoms with a ``2D adatom vapor" on the surface. Then,
$L_s$ is of order the
lattice spacing $a$, and Eq.~(\ref{N})
yields
\begin{equation}
w^2\mbox{\textit{\L}}^2\approx k {c_{eq}}\mbox{\textit{\L}} a t.
\end{equation}

Using Eq.~(\ref{e:wL3rd}) leads to $\mbox{\textit{\L}}^{5/3}\sim t$, and eventually to
\begin{equation}
\label{gad}
w\sim t^{1/5} \quad \rightarrow \quad G(t)\sim t^{2/5}.
\end{equation}
In comparison, $G(t)\sim
t^{1/2}$ for a straight step or an isolated 2D island \cite{NCBBrownian,pimpi}.

ii) For mass transport by step-edge diffusion along a
portion of step of size $\mbox{\textit{\L}}$, ${1}/{\tau^*}\approx D_e/\mbox{\textit{\L}}^2$, where
$D_e$ is the edge diffusion coefficient. Again in this case $L_s \approx a$, so that Eq.~(\ref{N})
becomes
\begin{equation}
N(t)\approx tc_{eq}D_ea/\mbox{\textit{\L}}.
\label{e:NtAD}
\end{equation}

From Eq.~(\ref{e:areafluc})
\begin{equation}
w^2\approx tc_{eq}D_e a/\mbox{\textit{\L}}^3.
\end{equation}
With Eq.~(\ref{e:wL3rd}) we now find $t\sim\mbox{\textit{\L}}^{11/3},$
so that
\begin{equation}
\label{e:gpdd}
w\approx t^{1/11} \quad \rightarrow \quad G(t)\sim t^{2/11}.
\end{equation}
\noindent ($\beta = 1/11$) for a crystal facet fluctuating through step edge diffusion. In comparison,
$G(t)\sim t^{1/4}$ ($\beta = 1/2$) for a straight
step or an isolated (large) 2D island \cite{NCBBrownian,pimpi}.

\subsection{Scaling in Curved Geometry with Asymmetric Potential}

We have also considered this problem from the perspective of scaling theory.
This approach begins by writing a Langevin equation
for the motion of
the edge of a crystal facet, \textcolor[rgb]{0.00,0.00,0.00}{or of an island}. In polar
coordinates, $r(\theta,t)$, the \textcolor[rgb]{0.00,0.00,0.00}{facet/island} radius (the
position of the
edge) in direction $\theta$ and at time $t$,
satisfies
the stochastic differential
equation
\begin{equation}
\frac{\partial r(\theta,t)}{\partial t}= f[r(\theta,t),r_{\theta},r_{\theta\theta},\ldots]  +\eta(\theta,t),
\end{equation}
\noindent where $r_\theta\equiv \partial r/\partial\theta$, \textcolor[rgb]{0.00,0.00,0.00}{$r_{\theta\theta}\equiv \partial^2 r/\partial\theta^2$, and so on for higher-order derivatives. The function $f$ describes the deterministic relaxation of
fluctuations and depends in principle on the facet radius and its higher-order derivatives with respect to $\theta$. Finally, $\eta(\theta,t)$ is a
white noise, which is conservative or non-conservative for SED or A/D, respectively.}

To obtain $f$  we assume
that the facet/island is delimited by
a closed step of free
energy per length $\beta(\theta)$.  Neglecting step-step
interactions (as in Ref.~\cite{ferrari04}),
the free energy of the facet/island is
\begin{equation}
F=\int_0^{2\pi}\! \beta(\vartheta)\sqrt{r^2+r_\theta^2}\; d\theta,
\end{equation}
where $\vartheta$ is the local direction of the step \cite{sk}.

Assuming for simplicity (as in Ref.~\cite{ferrari04}) an isotropic step free energy
$\beta(\theta)\! =\! \beta$ (implying a circular facet or island) one can compute
straightforwardly the
excess chemical potential with respect to a perfectly circular step edge,
which is given by the Gibbs-Thomson relation (see
\cite{sk,eins,pivi})
\begin{equation}
\delta\mu=\Omega \beta \left(\frac{r^2-rr_{\theta\theta}+2r_\theta^2}{\left(r^2+r_\theta^2
\right)^{3/2}}-\frac{1}{\rho_0}\right)
\end{equation}
where the first term in the parentheses is the step curvature and $\rho_0$ the average facet/island radius.

In order to study fluctuations around
$\rho_0$,
it is useful to introduce the new
variable
$\tilde r(\theta,t)= [r(\theta,t)-\rho_0]/\rho_0$. In terms of this
variable, the excess chemical potential
reads
\begin{equation}
\delta\mu=\frac{\Omega \beta }{\rho_0}\frac{(1+\tilde r)(1+\tilde r-\tilde
r_{\theta\theta})+2\tilde
r_\theta^2}{\left[(1+\tilde r)^2+\tilde r_\theta^2
\right]^{3/2}}-\frac{\Omega \beta }{\rho_0}.
\label{deltamu}
\end{equation}
Expanding around $\tilde r=0$, discarding all terms in $\tilde r$ as much smaller than unity but
keeping the
lowest nonlinear
terms in the derivative $\tilde r_\theta$, we obtain
\begin{equation}
\label{mu}
\delta\mu\approx \frac{\Omega \beta }{\rho_0}\left(-\tilde
r_{\theta\theta}+\frac{1}{2}\tilde r_\theta^2
\right).
\end{equation}

With the chemical potential in hand, we can model the step-edge fluctuations as a Langevin
equation for the two limiting cases considered previously.  (\textcolor[rgb]{0.00,0.00,0.00}{In experiments on  real crystals, of course, one can observe either of these two limiting behaviors or both, as well as crossover between them, depending on observational time scales and temperature. In metals, for instance, SED is known to dominate over A/D at low temperature, when thermal energy is not large enough to allow atoms to detach from a step edge. Therefore, A/D will only become observable at long times, with temperature determining how long a ``long time" is.})
i) In A/D  atoms ``evaporate from'' and ``condense
into'' the step edge. Accordingly,
we write \cite{P-SSL,eins,pivi}
\begin{equation}
\label{AD}
{\partial \tilde r(\theta,t)\over\partial t}=-{\Gamma_{AD}\over
k_BT}\delta\mu(\tilde r,\tilde r_\theta,
\tilde
r_{\theta\theta}) + \eta(\theta,t)
\end{equation}
where $\Gamma_{AD}$ is the attachment-detachment kinetic coefficient,
and $\eta(\theta,t)$ is a
Gaussian white noise.
Inserting Eq.~(\ref{mu}) into Eq.~(\ref{AD}) we find
\begin{equation}
\label{AD1}
{\partial \tilde r(\theta,t)\over\partial t}={\Gamma_{AD}\over k_BT}
{\Omega \beta \over\rho_0^3}\left
[{\partial^2\tilde
r\over\partial\theta^2}-{1\over2}\left({\partial\tilde
r\over\partial\theta}\right)^2\right] + \eta
(\theta,t).
\end{equation}

ii) Similarly for SED, we use conserved dynamics to represent atomic
diffusion along the step edge.
Accordingly, we write \cite{P-SSL,eins,pivi}
\begin{equation}
\label{SED}
{\partial \tilde r(\theta,t)\over\partial t}=\frac{\Gamma_{SED}}{
k_BT\rho_0^2}{\partial^2\over\partial\theta^2}
\delta\mu(\tilde
r,\tilde r_\theta,\tilde r_{\theta\theta}) + \eta_C(\theta,t)
\end{equation}
where $\Gamma_{SED}$ is the step-edge -diffusion kinetic coefficient,
and $\eta_C(\theta,t)$ is a conserved Gaussian white noise \textcolor[rgb]{0.00,0.00,0.00}{\cite{noise}}.
Inserting Eq.~(\ref{mu}) into Eq.~(\ref{SED}),  we get
\begin{equation}
\label{SED1}
{\partial \tilde r(\theta,t)\over\partial t}=\frac{\Gamma_{SED}}{k_BT}\frac{ \Omega \beta}{\rho_0\textcolor[rgb]{0.00,0.00,0.00}{^3}}
\left[-{\partial^4\tilde
r\over\partial\theta^4}+{1\over2}{\partial^2\over\partial\theta^2}\left({\partial\tilde
r\over\partial
\theta}\right)^2\right]
+ \eta_C(\theta,t).
\end{equation}

Eqs.~(\ref{AD1}) and (\ref{SED1}) resemble the Kardar-Parisi-Zhang (KPZ)
\cite{KPZ} equation and its conserved counterpart (the ``Montreal model") \cite{mont},
respectively. Here, however, the non-linear terms come from the equilibrium curvature of the
interface, while they are induced by non-equilibrium effects in the KPZ and
Montreal models. \textcolor[rgb]{0.00,0.00,0.00}{A closed (curved) step, such as the border of a facet or an island, has an ``inside" and an ``outside". The chemical potential in Eq.~(\ref{deltamu}) discriminates between inward and outward fluctuations of the step radius around its mean value $\rho_0$, thus breaking the radial symmetry. This is the origin of the ``asymmetric" KPZ-like nonlinearity in Eqs.~(\ref{AD1}) and (\ref{SED1}), which indeed breaks the $\tilde r \to - \tilde r$ symmetry of the equations. The asymmetry is insufficient to change the universality class of the fluctuating step.} We will
see that KPZ or Montreal exponents (see, e.g., Ref.~\cite{bs}) are \textcolor[rgb]{0.00,0.00,0.00}{still} expected for
the fluctuations of the edge of a small-enough (viz., smaller than the capillary length $k_BT/\beta$) island. In contrast, a facet should exhibit the exponents that we have computed in the previous subsection, which are neither KPZ nor Montreal. \textcolor[rgb]{0.00,0.00,0.00}{The difference stems from the interactions of the fact edge with the neighboring steps, which limit the amplitude of the fluctuations and change the character of the noise term in Eqs.~(\ref{AD1}) and (\ref{SED1}). In the previous subsection, the effect of interactions was \textit{implicitly} introduced through the assumption that the exponent $\alpha=1/3$. In the following, we derive the latter, too, using a scaling argument. Step-step interactions constitute another source of asymmetry at a facet edge, since steps are only present on one side of the fluctuating step. We will investigate the interaction-induced asymmetry in Section 5.}

To illustrate how our scaling arguments work, we first consider fluctuations of a
straight step in the A/D case.  \textcolor[rgb]{0.00,0.00,0.00}{(The rescaling in angle assumes that the angle is small, so that it essentially parametrizes distance along the arc.)} The step-edge fluctuations (in the A/D case) obey the linear equation \cite{eins,pivi}
\begin{equation}
\label{AD2}
{\partial x(y,t)\over\partial t}={\Gamma_{AD }\Omega \beta\over k_BT} {
}{\partial^2x\over\partial y^2} +
\eta(y,t).
\end{equation}

Rather than simply solving this linear equation, we use it to
illustrate the scaling argument.
Assume that
the linear size $\mbox{\textit{\L}}$ along the step edge is dilated by a factor $b$,
$\mbox{\textit{\L}}'=b\mbox{\textit{\L}}$ (where primed variables
denote rescaled quantities). Scaling implies that the width
$w$ of a fluctuation varies as $w\sim \mbox{\textit{\L}}^{\alpha}$, so that
$w'=b^{\alpha}w$. (See, e.g., Ref.~\cite{bs}.)  The typical time
needed to develop a
fluctuation of size $\mbox{\textit{\L}}$ scales as $t\sim \mbox{\textit{\L}}^{z}$, so that $t'=b^zt$.
The time derivative in Eq.~(\ref{AD2})
scales then as
\begin{equation}
\label{scal1}
{\partial x'(y',t')\over\partial t'}= b^{\alpha-z}\;{\partial
x(y,t)\over\partial t},
\end{equation}
\noindent while the Laplacian term scales as
\begin{equation}
\label{scal2}
{\partial^2  x'(y', t')\over\partial y'^2}= b^{\alpha-2}\;{\partial^2
x(y,t)\over\partial y^2}.
\end{equation}
\noindent Then equating the scaling exponents of $b$ in Eqs.~(\ref{scal1}) and (\ref{scal2}) yields $z=2$.

The scaling
exponent $\alpha$ depends on the
scaling behavior of
the noise term in the problem of interest.
If the step is isolated, the step edge should be treated as a 1D
interface. Then, since $\eta(y_1,t_1)\eta(y_2,t_2) = \delta(y_1-y_2)\delta(t_1-t_2)$,
the noise term scales as
\begin{equation}
\label{noise1}
{ \eta'(y',t')}= b^{-(1+z)/2}\;{\eta(y,t)}.
\end{equation}

Equating the scaling exponents of $b$ in Eqs.~(\ref{scal1}) and (\ref{noise1}) and using $z=2$ gives the random-walk value
\begin{equation}
\label{e:rdwalk}
\alpha=1/2.
\end{equation}
\noindent These results are tabulated in the first line of
Table \ref{tb:scale}.  The second line gives analogous results for the SED case.

If the step is inside a train, as on a vicinal surface, then \textcolor[rgb]{0.00,0.00,0.00}{Eq.~(\ref{AD2}) must be replaced by a similar second-order equation for the surface profile $ z=z(x,y)$. With $x$ and $y$ chosen so that they are perpendicular and parallel to the average step direction, respectively, the Laplacian in Eq.~(\ref{AD2}) is replaced by second derivatives of $z$ with respect to $x$ and $y$ with different coefficients, reflecting that the stiffness of a vicinal surface differs in directions parallel and perpendicular to the steps. The scaling of these terms, however, does not change with respect to Eq.~(\ref{scal2}). The important point is that in this case the fluctuations take on a 2D character, so that} the
noise term scales as
\begin{equation}
\label{noise2}
{ \eta'(x', y',t')}= b^{-(2+z)/2}\;{\eta(x,y,t)}.
\end{equation}
\noindent Equating the scaling exponents of $b$ in Eqs.~(\ref{scal1}) and (\ref{noise1}) and using $z=2$ now yields
$\alpha=0$, corresponding to the expected logarithmic scaling \cite{role} $w\sim \ln \mbox{\textit{\L}}$.

\begin{table}[b]
%\begin{ruledtabular}
\begin{tabular}{|l||c|c|c|c||c|c|}
\hline
Class & $\partial/\partial t$  & L $\nabla^{2,4}$& NL KPZ & Noise & \quad$\alpha$\quad & $z$ \\ \hline
Isolated A/D & $\alpha \! -\! z$ & $\alpha \! -\! 2$ & -- & $-(1 \! +\! z)/2$ &  1/2  & 2   \\ \hline
Isolated SED & $\alpha \! -\! z$ & $\alpha \! -\! 4$ & -- & $-(3 \! +\! z)/2$ &  1/2  & 4   \\ \hline
Train A/D & $\alpha \! -\! z$ & $\alpha \! -\! 2$ & -- & $-(2 \! +\! z)/2$ &  0 (ln)  & 2   \\ \hline \hline
Asym A/D & $\alpha \! -\! z$ & $\alpha \! -\! 2$ & $2\alpha \! -\! 2$ & $-(1 \! +\! z)/2$ &  1/3  & 5/3   \\ \hline
Asym SED & $\alpha \! -\! z$ & $\alpha \! -\! 4$ & $2\alpha \! -\! 4$ & $-(3 \! +\! z)/2$ &  1/3  & 11/3   \\ \hline
\end{tabular}
\caption[shrtab]{Summary of exponents resulting from scaling arguments for the evolution, [linear] relaxation, and noise terms of the relevant Langevin equations, as well as the nonlinear KPZ term for facet edges.
The deduced values of the roughness exponent $\alpha$ and the dynamic exponent $z$ are then listed.  See text for details.}
\label{tb:scale}
\end{table}

We turn now to the topic of primary concern, the scaling behavior of nonlinear Eqs.~(\ref{AD1})
and (\ref{SED1}) for a facet edge, for which
the nonlinearity dominates the scaling.
This nonlinearity comes from the curvature of the
step edge (Gibbs-Thomson effect), but \textcolor[rgb]{0.00,0.00,0.00}{the fluctuation spectrum may differ} differ in a subtle way from the analogous term in the KPZ equation. \textcolor[rgb]{0.00,0.00,0.00}{The latter is an equation for a growing interface, which roughens during growth. Because of the nonlinear term, the noise-induced roughening of the interface cannot be captured by a simple power-counting argument such as we used for the linear Eq.~(\ref{AD2}). Indeed, except for the 1D case, the scaling of the KPZ equations is still an open problem. Here, we are addressing equilibrium fluctuations of the interface; we can expect that their scaling will differ from the growth case. In fact, the different physical situations represented by an island edge and by a facet shoreline suggest that the noise term has to be treated differently in the two cases.  Ultimately,} the difference stems from the fact that, unlike the boundary step of a facet, an island edge is free to fluctuate, the amplitude $w$ of its fluctuations being limited only by the size of the island.
Because of the hindrance of neighboring steps, the fluctuations of a facet are constrained to smaller amplitudes than those of an island of comparable size.  As noted above, the non-linear term becomes important only for small (relative to the capillary length) islands. However,
the radius of an island has to be larger than a minimum value in order for
the island to be stable \cite{Krishnamachari}.
More details can be found elsewhere \cite{DSPEW,DW,Dphd}.

The main conclusion is that the step edge bordering an island may have
larger-amplitude fluctuations than does
the facet edge, since
the latter is limited by the presence of the neighboring steps. (Likewise, since these steps are only on one side, the facet edge has larger fluctuations than a step in the middle of a step train.) Thus, the noise terms scale
differently for a facet and for an island, leading to different
temporal and spatial scaling behaviors. Again $w$ is the
width of a step-edge protrusion of size \textit{\L}. \textcolor[rgb]{0.00,0.00,0.00}{In order to proceed as simply as possible, we decided to resort to the approach} in
Hentschel and Family \cite{hf}, \textcolor[rgb]{0.00,0.00,0.00}{which was shown to yield the exact scaling in 1D and a very good approximation in higher dimensions. In this approach, one takes} the length $S$ of a \textcolor[rgb]{0.00,0.00,0.00}{step fluctuation of amplitude $w$  to vary} as $S \approx (w^2+\mbox{\textit{\L}}^2)^{1/2} \sim $  \textit{\L} or $w$ for amplitude fluctuations that are small or large, respectively.  Assuming that atoms are added (or
removed) randomly to the step edge (for either A/D or SED), the relative fluctuations of the length of
the edge are just $\Delta S/S \approx S^{-1/2}$, indicating that it is reasonable to assume that the noise term in our stochastic
equations scales as
\textcolor[rgb]{0.00,0.00,0.00}{
\begin{equation}
\label{EF}
\eta(S,t)\sim (St)^{-1/2} .
\end{equation}
As a consequence, for fluctuations of small amplitude we have
\begin{equation}
\label{EF1}
\eta(S',t')\sim (\textit{\L}'t')^{-1/2}\sim b^{-(\textcolor[rgb]{0.00,0.00,0.00}{1}+z)/2} \eta(S,t),
\end{equation}
while for large amplitude
\begin{equation}
\label{EF2}
\eta(S',t')\sim (w' t')^{-1/2}\sim b^{-(\alpha+z)/2} \eta(S,t).
\end{equation}
}
\textcolor[rgb]{0.00,0.00,0.00}{The last comment concerns the scaling variable. In Eqs.~(\ref{AD1}) and (\ref{SED1}) the variable is an angle, but since the facet (island) radius $\rho_0$ is kept fixed, the quantity $\rho_0\theta$  measures the length along the step edge, and scales as a length. We can now address the scaling.} Consider first a facet fluctuating by attachment-detachment,
Eq.~(\ref{AD}). In this case, step
fluctuations are limited
in amplitude by neighboring steps. \textcolor[rgb]{0.00,0.00,0.00}{ Equating the scaling of the time derivative to that of the noise term in Eq.~(\ref{EF1}) yields}
\begin{equation}
\label{tdnt}
z=2\alpha+1 .
\end{equation}

\noindent The nonlinear term  $\tilde r_\theta^2$ scales as
\begin{equation}
\label{nl}
\left({\partial \tilde
r'\over\partial\theta'}\right)^2=b^{2\alpha-2}\left({\partial \tilde
r\over\partial\theta}\right)^2
\end{equation}

\noindent Equating Eq.~(\ref{nl}) to the noise term Eq.~(\ref{EF1}) yields
\begin{equation}
\label{sdnt}
4\alpha+z=3
\end{equation}

\noindent From Eqs.~(\ref{tdnt}) and (\ref{sdnt}) we finally get FPS's result
\begin{equation}
\alpha=1/3 \qquad \Rightarrow \qquad \tilde r\sim \mbox{\textit{\L}}^{1/3} .
\label{fea}
\end{equation}

The dynamic scaling of step fluctuations turns out to be what we
computed previously: From Eqs.~(\ref{tdnt}) and
(\ref{fea}), e.g., we obtain
\begin{equation}
\label{time}
z = 5/3 \qquad \Rightarrow \qquad \alpha/z=\beta={1/5} ,
\end{equation}
which, recalling that $G(t)\sim t^{2\beta}$, coincides with
Eq.~(\ref{gad}).

Facet fluctuations driven by step-edge diffusion obey Eq.~(\ref{SED1}).
The conserved noise term scales
as
\begin{equation}
\label{connoisefacet}
{ \eta_C'(S',t')}= b^{-(3+z)/2}\;{\eta_C(S,t)}.
\end{equation}

\noindent The conserved nonlinear term scales now as
\begin{equation}
\label{cnl}
{\partial^2\over\partial\theta'^2}\left({\partial \tilde
r'\over\partial\theta'}\right)2=b^{2\alpha-4}{\partial^2\over\partial\theta^2}\left(\frac{\partial
\tilde r}{\partial\theta}\right)^2
\end{equation}

\noindent Equating (\ref{connoisefacet}) and (\ref{cnl}) yields
\begin{equation}
\label{sd11}
4\alpha+z=5
\end{equation}

\noindent Equating the time derivative to the noise term Eq.~(\ref{connoisefacet})
yields
\begin{equation}
\label{tdnt11}
2\alpha=z-3 .
\end{equation}

\noindent Together Eqs.~(\ref{sd11}) and (\ref{tdnt11}) yield
\begin{equation}
%\label{ncpl2}
\alpha=1/3, \;\;\; z =11/3 \quad \Rightarrow \quad \beta=1/11
\label{ncpl3}
\end{equation}
as in Eq.~(\ref{e:gpdd}).

As mentioned above, KPZ-like or Montreal-like exponents are expected to appear in the
fluctuations of a small island edge, for non-conserved and conserved dynamics,
respectively. For the latter, fluctuations are not hindered, and the noise scales as in Eq.~(\ref{EF1}).
It then follows straightforwardly that the scaling relations
$\alpha+z=2$ and $3\alpha=z$,
implying $\alpha=1/2$
as for a random walk [cf. Eq.~(\ref{e:rdwalk})], as well as the KPZ result (first noted in Ref.~\cite{P-SSL}) that
\begin{equation}
\beta=1/3.
\end{equation}
This result clearly
applies only to small islands with large curvature;
otherwise, the scaling should be that of a straight step.

Table \ref{tb:ucls} summarizes the various universality classes \cite{bs,hf} that can arise for different types of spatial confinement for non-conserved and conserved kinetics .

\begin{table}[t]
%\begin{ruledtabular}
\begin{tabular}{l||l|c|c|c||l|c|c|c}
Geom. & A/D  & \quad$\alpha$\quad &\quad $2\beta$ \quad& \quad $z$ \quad& SED  &\quad $\alpha$ \quad & \quad $2\beta$ \quad & \quad $z$ \quad  \\ \hline
Free & $l$M$^2_1$(EW) & $\frac{1}{2}$ & $\frac{1}{2}$ & 2 & $l$C$^4_1$ & $\frac{1}{2}$ & $\frac{1}{4}$ & 4 \\
Sym-cfn & $l$M$^2_2$ & 0 & 0 & 2 & $l$C$^4_2$ & 0 & 0 & 4 \\
Asy-cfn & KYP\! \cite{kyp}& $\frac{1}{3}$ & $\frac{2}{5}$ & $\frac{5}{3}$ & $n$C$^4_1$ & $\frac{1}{3}$ & $\frac{2}{11}$ & $\frac{11}{3}$  \\ \hline
KPZ & $n$N$^2_1$ & $\frac{1}{2}$ & $\frac{2}{3}$ & $\frac{3}{2}$ & $n$M$^4_2$ & $\frac{2}{3}$ & $\frac{2}{5}$ & $\frac{10}{3}$
\end{tabular}
\caption[shrtab]{ Summary of the dynamical scaling universality classes for crystallite steps.  The geometries included are:  Free = an isolated step or island edge, Sym-cfn = steps symmetrically confined by the nearby steps as in a step bunch, and Asy-cfn =  steps confined by an asymmetric potential, esp.\ a facet edge.  The KPZ class is included for comparison.  In the underlying Langevin equation (cf.\ Ref.~\cite{bs}), $l$ or $n$ indicates whether the equation is linear or non-linear (has a KPZ term).  C or N indicates whether the deterministic part and the noise are conservative or non-conservative; M denotes mixed, with the former conservative but the noise not.  The superscript (2 or 4) indicates the power of $\nabla$ in the linear conservative term, while the subscript gives the dimensionality of the independent variable. From Ref.~\cite{D-PRL}.
}
\label{tb:ucls}
\end{table}

Finite-volume effects on \textcolor[rgb]{0.00,0.00,0.00}{supported} nano-crystallites with a Gruber-Mullins-Pokrovsky-Talapov surface free energy density \cite{GMPT} have been found to produce metastable states with different crystal shapes \cite{UN89} for a given crystal-substrate interface boundary condition \cite{8,12}.  All shapes have a facet smoothly connected to a vicinal region, which obeys an $x^{3/2}$ shape power law in equilibrium \cite{13}.
Once a crystallite attains a stable state, the step that serves as the interface between the facet and the vicinal region (see Fig.~1b)) fluctuates around its stable position, which is determined by the asymmetric potential established by step-step interactions and
the ``reservoir" chemical potential of the crystallite.

\section{STM Experiments}

\begin{figure}[t]
\includegraphics[width=8.2 cm]{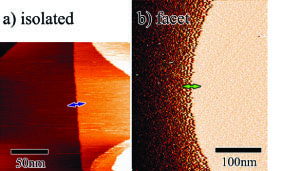}
\caption {An STM image of a) an isolated step on a crystallite facet (room temperature) and b) a crystal facet edge (350K). The small superimposed double-arrow indicates the tip path that leads to line-scan images as in Fig.~\ref{f:linescan}. From Refs.~\cite{D-PRL,Dphd}.
}
{\label{f:stm}}
\end{figure}

\begin{figure}[t]
\includegraphics[width=8 cm,height=5cm]{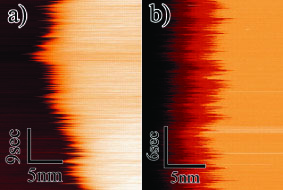}
\caption {Segment of a line-scan pseudoimage of a) an isolated step (step from screw dislocation) and b) a facet-edge at 350K, showing also the correlated fluctuations of the neighboring steps.  The time interval between lines is 0.02 s, and 2000 lines are measured per image. From Refs.~\cite{D-PRL,Dphd}.
}
{\label{f:linescan}}
\end{figure}

The first experimental observations of the novel scaling predicted for facet-edge fluctuations on crystallites were performed at the University of Maryland by Masashi Degawa working under Ellen D.\ Williams \cite{D-PRL}.   Crystallites were formed by depositing a 20--30 nm Pb film at room temperature on a Ru(0001) substrate in UHV \cite{15}, and subsequently dewetting at 620 K. The liquid Pb droplets solidified upon slow cooling and were left to equilibrate to a stable state at the $T$ of the experiment \cite{15,Thurmer03,Thurmer01}.  The crystallites were observed with a variable-temperature scanning tunneling microscope (VT-STM) after equilibration.  Figure~\ref{f:stm} depicts an STM image of a) an isolated step (at room temperature) and b) facet-edge (at 350K).  A crystallite in a stable state as shown in b) has a flat, close-to-circular (111) facet and a smoothly-connecting vicinal region.

\begin{figure}[t]
\includegraphics[width=8 cm]{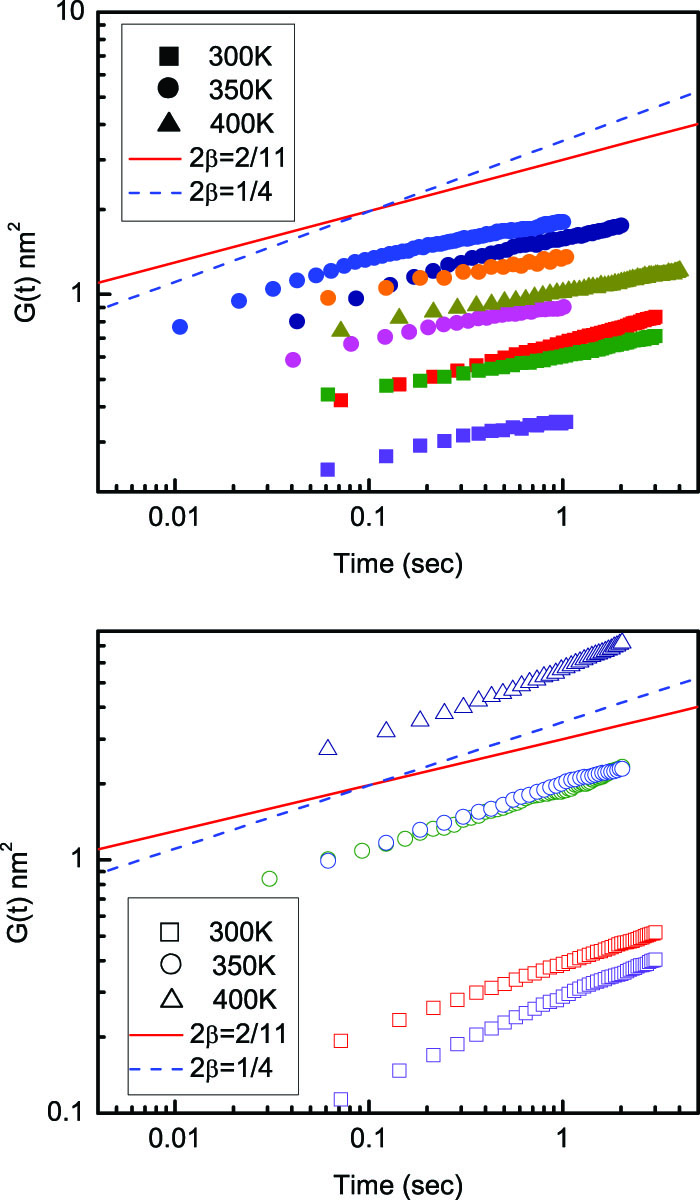}
\caption {Log-log plot of $G([y_0,]t)$ (cf.\ Eq.~(\ref{e:Gt})) for a) facet edges and b) isolated steps with facet radii from 60 to 190 nm.  The symbols represent: 300K (squares), 350K (circles), and 400K (triangles). For guidance, solid and dashed lines show slopes \textcolor[rgb]{0.00,0.00,0.00}{$2/11 = 0.\overline{18}$} and 1/4, respectively.  Individual fits to each of the data sets yield slopes of a) facet edges:  300K: 0.18(1), 0.13(6), 0.13(2); 350K: 0.17(4), 0.17(4), 0.12(3), 0.11(5); 400K: 0.12(12), and b) isolated steps: 300K: 0.32(3), 0.26(1); 350K: 0.24(3), 0.24(4); 400K: 0.30(4). From Refs.~\cite{D-PRL,Dphd}.
}
{\label{f:log}}
\end{figure}

Since STM images are scans rather than instantaneous ``snapshots", the data for dynamic scaling can be more compelling than that for static scaling.
By repeatedly scanning perpendicularly to a single position ($y_0$) along the facet-edge or step (cf.\ Fig.~\ref{f:stm}), we obtain a line-scan STM image \cite{giesen01} $x(t)$, as shown in Fig.~\ref{f:linescan} for a) an isolated step (step from a screw dislocation) and b) a facet-edge, both at 350K.  Digitized step displacement-positions $x(t)$ extracted from these ``pseudoimages" are used for statistical analysis.
To evaluate the growth exponent $\beta$, we calculated the early behavior of the time correlation function $G(t)$ given in Eq.~(\ref{e:Gt}).  To evaluate the roughness exponent $\alpha$, we can calculate either the saturation value of the width $w$ of the fluctuating step as in Eq.~(\ref{e:wL}) or the spatial correlation function $G(y,t_0) \sim y^{2\alpha}$ for $y$ less than the correlation length \cite{role}.

Figure~\ref{f:log} shows $G(t)$ determined for a) facet-edges and b) isolated step-edges.  Squares, circles, and triangles represent measurements at 300K, 350K and 400K, respectively.  Each curve displays the average over the correlation functions for 10-30 measurements of $x(y_0,t)$.  The exponent 2$\beta$ for each temperature is obtained from the slope of the curve on the log-log plot; the values of these slopes are listed in the figure caption.  As expected the exponents show no systematic thermal dependence: from all data sets, the $\sigma^{-2}$-weighted average exponent is 2$\beta = 0.149 \pm 0.032$ for facet-edges and 2$\beta = 0.262\pm 0.021$ for isolated steps.  With over 99.9\% confidence (using Student's t-test), these values come from different parent populations.
Each of the two results is within one standard deviation, $\sigma$, of their respective predicted values of 2/11 and 1/4.

\begin{figure}[t]
\includegraphics[width=7 cm]{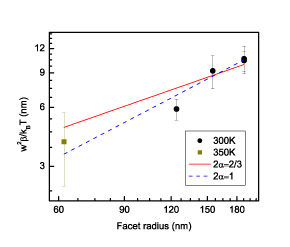}
\caption {Product of squared saturation width and reduced stiffness as a function of facet radius (facet-edge only).  Circles and squares are room temperature and 350K, respectively.  Solid and dashed lines are a fit to the 350K data with $\alpha$=1/3 and $\alpha$=1/2, respectively. From Refs.~\cite{D-PRL,Dphd}.
}
{\label{f:sat}}
\end{figure}

Determination of the roughness exponent $\alpha$ requires evaluation of the system-size dependence.  A detailed examination is a challenge beyond the capability of the STM experiments being used.  However, we can demonstrate that size does affect fluctuations.  Under the assumption $\mbox{\textit{\L}} \sim R$ (the facet radius) for confined steps (Fig.~\ref{f:log}a), we expect $w^2 \sim R^{2\alpha}$.  (For the unconfined steps, the system size is larger than the limitations imposed by the finite measurement time \cite{6}.)
The effect of the facet size is apparent in Fig.~\ref{f:log}a since the three upper sets of data at 350K were taken on larger crystallites, having radii greater than 100 nm. More quantitatively, Fig.~\ref{f:sat} plots the characteristic \textit{length} $w^2\tilde\beta/k_BT$ vs.\ facet radius at 300K and 350K, using for  the step stiffness $\tilde\beta$ the values 0.339 eV/nm and 0.327 eV/nm \cite{17}, respectively.   Fits to the data yield exponents within the predicted range of $\alpha$=1/3 (solid) to $\alpha$=1/2 (dash). Although there are insufficient data to distinguish between these two values \cite{notealph}, the results do show that $R$ influences the fluctuations, providing further evidence that effects of crystal confinement govern the behavior of $G(t)$.

%In Fig.~\ref{f:sat}, $w^2\tilde\beta /k_BT$ is plotted as a function of facet radius, where %$\tilde\beta$ is the step stiffness (= 0.339 eV/nm at room temperature and 0.327 eV/nm at
%80$^\circ$C \cite{17}).  The square and circles correspond to values at room temperature and %80$^\circ$C, respectively. Based on the assumption that the system size is proportional to
%the facet radius, the data are fit by $\alpha$ =1/3 (solid) and $\alpha$ =1/2 (dash) %\cite{notealph}.

The fluctuations of facet edges evidently belong to a different universality class of dynamic scaling from that of an isolated step on a surface.  In contrast to previous predictions for step exponents \cite{JW,giesen01,hiT,flynn}, this difference does not stem from the type of kinetics.  Instead, the effect is predicted to result from the coupling of the step chemical potential to the fluctuations:
For facet-edge fluctuations the step confinement is produced by an increase in local step chemical potential $\mu(x)$ when the step is displaced from equilibrium.  The functional behavior of $\mu(x)$ results from a competition between the step-repulsions from the vicinal region and the 2D pressure of the adatom density on the facet, which in turn stems from the constraints governing the crystallite shape \cite{8,UN89}. For a step symmetrically confined on a vicinal surface, the confinement corresponds to a potential that is quadratic in displacement \cite{bew}.  However, for the facet-edge step, the asymmetry in the $\mu(x)$ corresponds to an asymmetric confining potential that includes a cubic term in displacement \cite{19} and, consequently, leads to non-linear terms in the equation of motion discussed above.

\begin{figure}[h]
\includegraphics[width=9 cm]{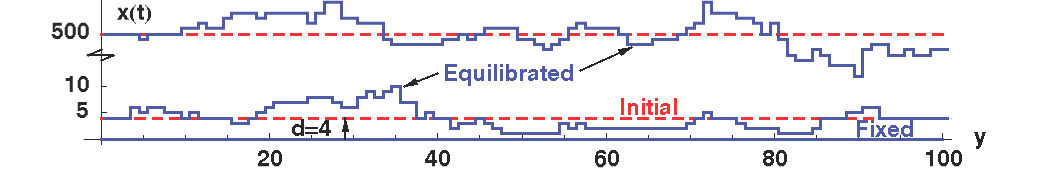}
\caption {Snapshot of the MC configuration, $x(y)$ of an isolated step (d=500) and d=4 for $T= \epsilon/k_B$ and $L_y$=100.  The initial step and the equilibrated step are shown as dashed and solid lines, respectively.  The line $x=0$ is the position of the fixed neighboring step. From Ref.~\cite{DSPEW}.
}
{\label{f:MCconf}}
\end{figure}

\begin{figure}[h]
\includegraphics[width=8 cm]{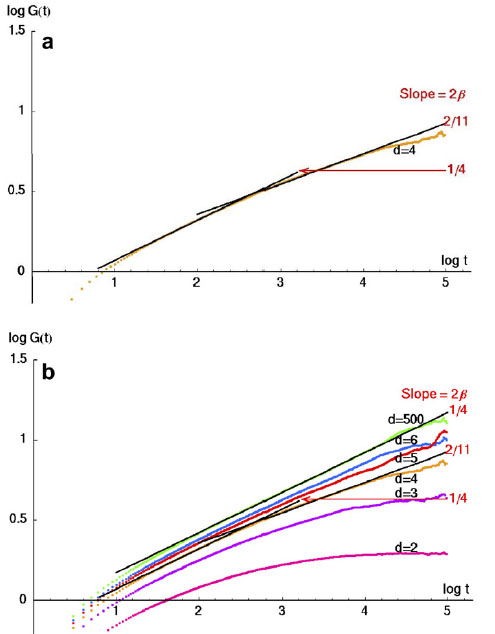}
\caption {Log-log plot of $G(y_0,t)$ obtained from $10^7$ MC steps per site after equilibration and averaged over 10 realizations.  In panel a) results for d=4 are plotted; the slope is \textcolor[rgb]{0.00,0.00,0.00}{clearly consistent with rather rapid crossover between the predicted values (lines included for convenience) $2\beta$=1/4 at small $t$ and then $2\beta$=2/11.}  In panel b) are results for values of $d$ between 2 and 500.  For $d$=2  logarithmic behavior ($\beta$=0) is observed, while for large values of $d$ ($d$=500) $2\beta$=1/4. Adapted from Refs.~\cite{DSPEW,Dphd}.
}
{\label{f:lnlntime}}
\end{figure}

\begin{figure}[h]
\includegraphics[width=8 cm]{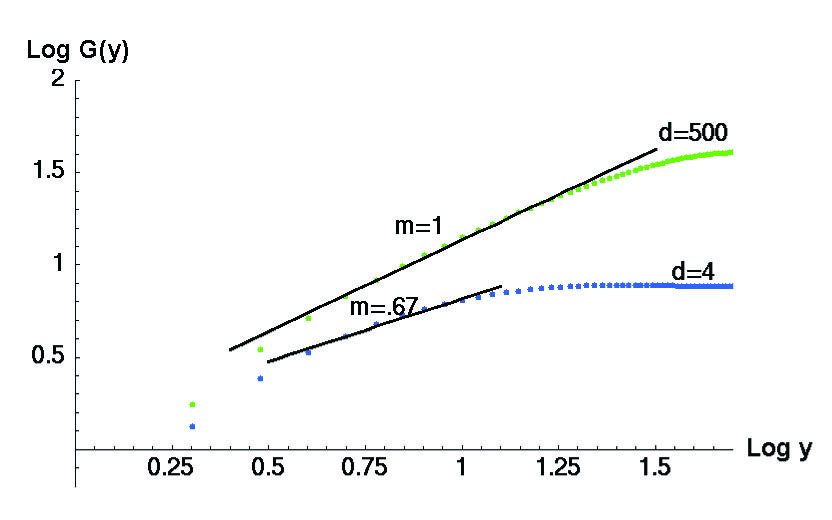}
\caption {Log-log plot of $G(y,t_0)$ obtained after equilibration for $d=4$ and $d=500$.  The fit of the slope $m$ (i.e. $2\alpha$)---over the linear regime where $y$ is less than the correlation length---is ~0.67 and ~1 for $d=4$ and $d=500$, respectively. From Refs.~\cite{DSPEW,Dphd}.
}
{\label{f:lnlnspace}}
\end{figure}

\section{Monte Carlo Simulations}

\textcolor[rgb]{0.00,0.00,0.00}{As discussed above, both geometry and interstep interactions break the front-back symmetry of the potential felt by a fluctuating step bordering a facet. Taking both effects into account is not feasible in a simple way. In Section 3 we accounted for the geometry by computing the chemical potential for a closed (circular) step. In this Section we consider a simpler geometry, in which steps are straight, and we consider the effect of the contact repulsion (entropic interaction) between neighboring steps. This will help us} to elucidate the \textcolor[rgb]{0.00,0.00,0.00} {role played by step-step interactions in determining the} physics of asymmetric confinement in a conserved-volume system. \textcolor[rgb]{0.00,0.00,0.00}{To this end, we assumed SED-limited kinetics, and} we carried out standard Monte Carlo (MC) simulations of a ``toy" TSK model on a square lattice in which a single ``active" step is placed a distance $d$ lattice constants \textcolor[rgb]{0.00,0.00,0.00}{(in the $\hat{x}$ direction)} from a second fixed straight step \textcolor[rgb]{0.00,0.00,0.00}{(i.e. a 1+1D SOS model with this and other restrictions)}; both steps have projected length $L_y$ \cite{Dphd}.  For convenience we set $k_BT$ to the energy $\epsilon$ of a unit length of step and assume only entropic repulsive interactions between the two steps.  The active step evolves by Kawasaki dynamics, with trial moves by ``atoms" at the step to neighboring sites along the step.  \textcolor[rgb]{0.00,0.00,0.00}{(In SOS language, neighboring sites increase and decrease by one unit.)}  Most of our runs were done with $L_y = 100$, with $\sim \! 10^8$ MCS; consistent with the high value of $z$ for SED dynamics, runs with $L_y = 200$ converge poorly.
Figure \ref{f:MCconf} shows a snapshot of a MC configuration of an essentially isolated step ($d=500 [a]$) and for $d=4$, both initially (dashed) and after equilibration (solid) \cite{D-PRL,DSPEW}.

To simulate confinement, the neighboring step is placed at distances $d$ = 2, 3, 4, 5, 6, and 500.  Figure \ref{f:lnlntime} shows G(t) obtained after equilibration
for $10^7$ MCS (MC steps per site) and averaged over
10 realizations.   Figure \ref{f:lnlntime}a shows results for just $d = 4$ for clarity.  The fit to the
data clearly shows crossover.  At early times the slope is
$m \sim 0.25$ while later $m \sim 0.18$, close to the predicted values
of 2$\beta =$ 1/4 and 2$\beta =$ 2/11 for isolated and facet (confined)
step-edge fluctuations, respectively, with SED kinetics.  In other words, once the step meanders enough to ``become aware" of the fixed step, $G(y,t_0)$ crosses over from random walk behavior to asymmetric conserved-volume confinement, then eventually begins to cross over to the flat late-time behavior of symmetrically confined steps.
Figure \ref{f:lnlntime}b adds to Figure \ref{f:lnlntime}a the results for $d$ = 2, 3, 5, 6, and 500. For
$d = 2$ the neighboring step is so close that no simple
power-law behavior is evident; passage from early to late behavior is quick, smooth, and broad. As the stepâstep distance
increases, the time until crossover from 2$\beta =$ 1/4 to
2$\beta =$ 2/11 also increases. When $d > 6$, crossover is not seen before the end of the
$10^7$ MCS, and the slope
remains 2$\beta =$ 1/4.
Thus, for $d \! = \! 6$ and especially for $d \! = \! 500$, much larger than the mean squared width of the step $w^2$, the fixed step never significantly influences the active one.
That the experimental value of 2$\beta \! = \! 0.15 \! \pm \! 0.03$ is somewhat below 2/11 weakly suggests (one-$\sigma$)  that some physical effect may be acting to reduce the growth exponent.
The possibility of extreme damping of fluctuations due to small step spacings, as for $d \! = \! 2$ in Fig.~\ref{f:lnlntime}b, is unlikely since the Pb measurements correspond to $d/w$ values $\sim$5--10, well above the strong-confinement regime \cite{D-PRL}.  Thorough analysis of a more detailed model would be needed for quantification.

Figure \ref{f:lnlnspace} shows results of $G(y,t_0)$ obtained
after equilibration for $d = 4$ and $d = 500$.
The initial linear portion of the log-log-plotted
data is fit to a slope $m \sim 0.67$ and $m \sim 1.0$ for $d = 4$
and $d = 500$, consistent with the prediction of $2\alpha = 2/3$
and $2\alpha = 1$ for confined and isolated steps, respectively \cite{Dphd,DSPEW}.
However, since the length of the step is limited to 100 and the linear scaling in Fig.~\ref{f:lnlnspace} is over less than a decade in $y$, these results are not fully conclusive.

As shown in Fig.~\ref{f:lnlntime}, after random-walk ($2\beta \! =\! 1/2$) evolution at the very outset (first few points), $G(t)$ quickly crosses over to isolated-step ($2\beta \! =\! 1/4$) behavior.  For $d \! = \! 4$,   For $d \! = \! 2$, confinement is so great that $G(t)$ progresses quickly from initial- to late-time evolution, with no clear intermediate regime.

\section{Conclusions}

The work presented here is a very good example of the interplay in statistical physics between exact results, scaling arguments, numerical simulations, and experiments. Spohn and coworkers have produced a novel, exact static result. This has motivated us to apply old scaling arguments to derive novel dynamical behaviors. In turn, the latter have opened new avenues  for experimentalists to explore. And the results of experiments have motivated numerical simulations of model systems. In particular, for the first time it has been possible to observe experimental evidence for a nonlinear term in equilibrium fluctuations. The result agrees with our predictions for the case of geometrically confined fluctuations.    When power-law temporal correlations are measured,  the measured value of the power $\beta$ is significantly smaller than the unconfined exponent of $\beta$=1/8, and is within 1$\sigma$ of the predicted value of $\beta$=1/11 for a universality class of dynamical scaling with $\alpha$=1/3 and $z$=11/3.  Thanks to the \textcolor[rgb]{0.00,0.00,0.00}{extensions by} Spohn and coworkers \textcolor[rgb]{0.00,0.00,0.00}{of earlier links between} KPZ behavior and the behavior of facet edges \cite{SBdN}, we were able, for the first time (to the best of our knowledge), to provide an example in which a KPZ--type equation of motion  accounted for equilibrium fluctuations. We were also able to experimentally verify the predictions of the theory.  The experiments spurred a more detailed numerical study of the problem. As a result, it was discovered that the fluctuations and equation of motion of steps at equilibrium are very sensitive to the step environment \cite{Tao}, a discovery that may introduce new opportunities for controlling the fabrication of nanostructures and for understanding new aspects of their dynamic properties.

\begin{acknowledgements}
Work at University of Maryland has been supported by the UMD-NSF MRSEC under grant DMR 05-20471; TLE is now supported partially by NSF-CHE 07-50334 and 13-05892.  Much of this paper is based on extensive collaboration with the experimental surface physics group at UMD, led by Ellen D. Williams  until 2010, with ongoing guidance by Janice Reutt-Robey and William G. Cullen, in particular with Masashi Degawa, whose dissertation research accounts for much of the content of this paper.  We also benefited from interactions with theory postdoc Ferenc Szalma and students Hailu Gebremarian and Timothy J. Stasevich.
\end{acknowledgements}


\begin{thebibliography}{99}
\bibitem{calA} In Ref.~\cite{ferrari04} $A$ is defined as $b_\infty^{\prime\prime}$; writing ${\cal A} = (1/2)A$ simplifies Eq.~(\ref{e:var2}).
\end{thebibliography}
\end{document}